\documentclass[amsmath, twocolumn, showpacs, aps]{revtex4-1}
\usepackage{bm}
\usepackage{graphics}
\usepackage{graphicx}
\usepackage{amssymb}
\usepackage{color}

\begin{document}

\title{Second order interference of chaotic light reflected from random medium}
\author{ A. Yu. Zyuzin}

\affiliation{ A.F. Ioffe Physico-Technical Institute of
Russian Academy of Sciences, 194021 St. Petersburg, Russia}


\begin{abstract}
We consider the reflection  from a random medium of light with short coherence length.
We  found that the second order correlation function of light can have a peak in a direction where the reflection angle is equal to angle of incidence. This occurs when the size of the region, from which light is collected, is larger than the coherence length.
\end{abstract}
\maketitle

\section{Introduction}
Considerable theoretical and experimental interest have been recently denoted in the field of strong scattering of quantum states of light.

It was theoretically proposed \cite{bib:been, bib:Lahini, bib:Lodahl1,bib:ott, bib:Cande} that the entanglement of light, i.e. the quantum nature of light, can be probed in the multi-photon scattering experiments \cite{bib:Lodahl2,bib:peeters, bib:Pires}.

Also of basic interest are questions of optical noise propagation in random medium \cite{bib:smolka}, 
and photon counting statistics  of multiple scattered light \cite{bib:balog}.

The reflection of light from multiple scattering medium on average is an angle independent and has a weak localization peak due to the constructive interference  in the backscattering direction. It is a precursor manifestation of Anderson localization. Pioneering work on weak localization of photon noise have been reported in \cite{bib:scalia}. We note, that the experiment was limited by a large light coherence length.

In this paper we consider a situation of arbitrary relation between the coherence length and system size.
We find that when the coherence length is smaller than the system size,  probability of two photon absorption develops a peak at reflection angle equal to incidence angle. The obtained peak constitutes a new characteristic of a light scattering from random medium.

\section{Definitions} 
 
We consider the light incident at direction $\textbf{n}$ on the surface of disordered medium and reflected after multiple scattering in direction $\textbf{m}$. 
Diffusion transport of light is characterized by the mean free path $l$, which is much smaller than the size of the medium.
\begin{figure}[t]  \centering
\includegraphics[width=5cm] {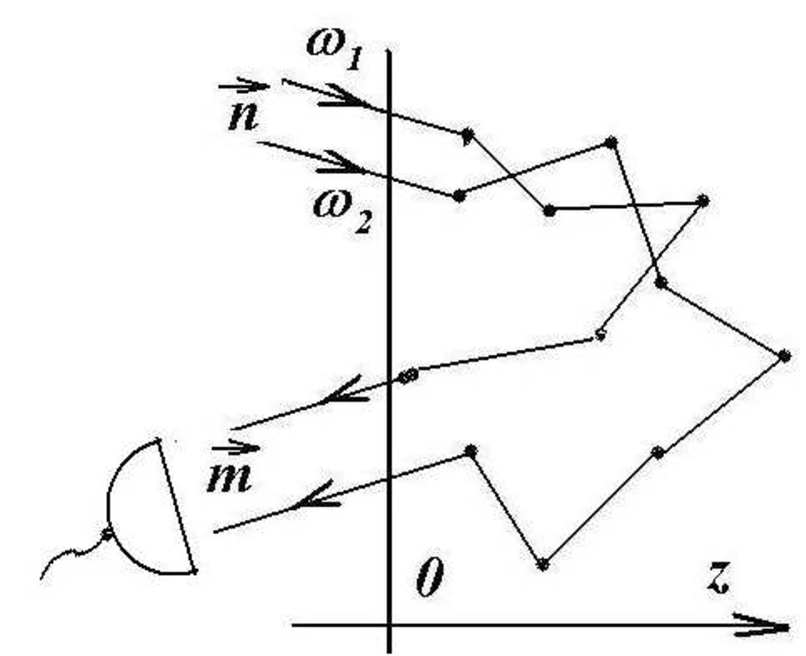}
\caption{Schematic picture of two-photon scattering in random medium, placed at $z>0$}\label{fig:1}
\end{figure}
The light reflected from the area of size $S$ in the direction $\textbf{m}$ is collected by the detector, as it is shown in figure \ref{fig:1}.

The probability of absorption of photons at points $x_{1}=(\textbf{r}_{1},t_{1})$ and $x_{2}=(\textbf{r}_{2},t_{2})$ of detector is defined by the second order correlation function
\begin{equation}
G^{(2)} = \mathrm{Sp}[\varrho A^{i}_{-}(x_{1})A^{j}_{-}(x_{2})A^{j}_{+}(x_{2})A^{i}_{+}(x_{1})],
\end{equation} 
where $ \textbf{A}_{+}(x)$ and $\textbf{A}_{-}(x)$ are positive and negative frequency parts of the vector fields, with $i,j=(x,y,z)$.

We assume that light is chaotic and characterized by the coherence length
comparable with the linear size of the area. The stationary Gaussian density operator $\varrho$ of incident light is defined by the
correlation function:
\begin{eqnarray}\label{cor-fun}
&~&\mathrm{Sp}[\varrho c^{+}_{\alpha _{1}}(t_1)
c^{+}_{\alpha _{2}}(t_2) 
c_{\beta_1}(t_2) 
c_{\beta_2}(t_1)]\\\nonumber
&=& 
\nu(\omega _{1}) \nu(\omega_{2})
(\delta_{\alpha_{1},\beta_{2}}\delta_{\alpha_{2} ,\beta_{1}}+\delta_{\alpha_{1},\beta_{1}}\delta_{\alpha_{2} ,\beta_{2}}\exp(i\omega_{12}t_{12})).
\end{eqnarray}
Here 
$c^{+}_{\alpha}$ and $c_{\alpha}$ are creation and annihillation operators of photons in state $\alpha \equiv \textbf{k},s$. 
$\textbf{k}|| \textbf{n}$ and $s=1,2$ are respectively the photon momentum and the polarization with complex polarization vector $\textbf{e}(\textbf{k},s)$. The difference between the frequencies of photon in states $\alpha_{1}$ and $\alpha_{2}$ is $\omega_{12}=\omega_{1}-\omega_{2}=c(k_{1}-k_{2})$.

In order to study the crossover from large to short coherence length of light compare to linear size of the area we 
assume that the spectral function $\nu(\omega)$ is Gaussian:
\begin{equation}
\nu(\omega)= N_{0} \exp[- \sigma(\omega-\Omega_{0})^2],
\end{equation} 

centered at $\Omega_{0}=ck_{0}$.

The coherence length of the chaotic light, which is characterized by the this spectral function, is 
\begin{equation}
 L_{c}=c\sqrt{\sigma}.
\end{equation}

We will calculate the disorder averaged value of the probability of absorption 
$\langle G^{(2)}\rangle$.  As a normalization constant we choose  the time independent intensity of light 
$\langle G^{(1)} \rangle = \langle \mathrm{Sp}[\varrho \textbf{A}_{-}(x)\textbf{A}_{+}(x)]\rangle$, averaged over disorder.
An expression for normalization constant will be given in IV section along with results.

Quantities, which will be calculated are 

\begin{equation}
 g^{(2)}(\textbf{n},\textbf{m},t_{12})\equiv\langle G^{(2)}\rangle/\langle G^{(1)} \rangle^{2}.
\end{equation} 

and Fourier transformation of $ g^{(2)}(\textbf{n},\textbf{m},t_{12})$ over $t_{12}$, which might be represented as 

\begin{equation}\label{four}
 g^{(2)}(\textbf{n},\textbf{m},\Omega)=2\pi  g^{(2)}_{0}(\textbf{n},\textbf{m})\delta(\Omega)+ g^{(2)}_{1}(\textbf{n},\textbf{m},\Omega) 
\end{equation} 

First and second terms here correspond to that in the definition (\ref{cor-fun}).

\section{Calculation of $ g^{(2)}(\textbf{n},\textbf{m},\Omega)$}
\begin{figure}[t]  \centering
\includegraphics[width=8cm] {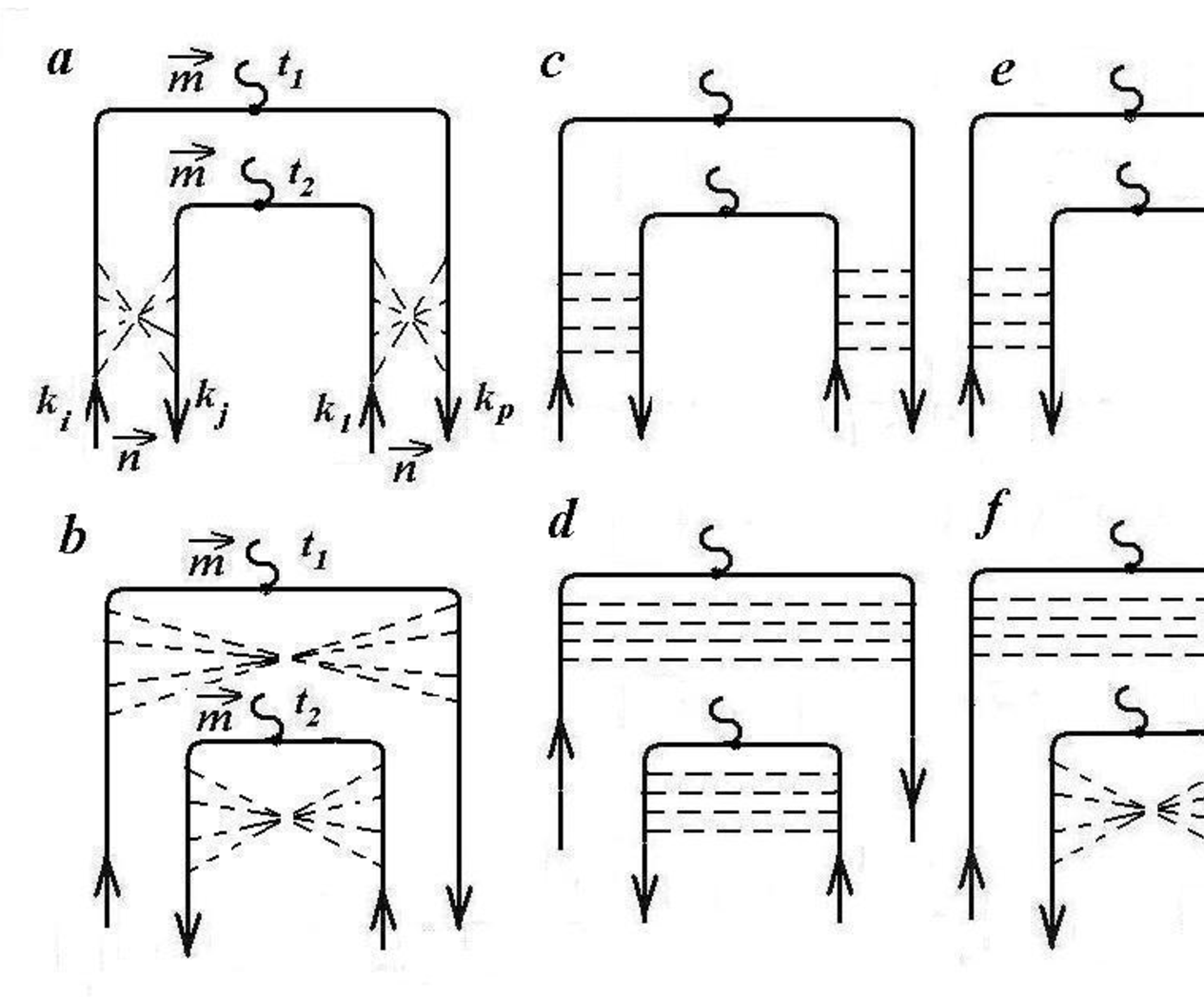}
\caption{Diagrams which contribute to the second order correlation function. The solid lines denote the light
Green's functions and the dashed lines represent scattering.}\label{fig:2}
\end{figure}

Diagrams describing Cooperon, Diffuson, and mixed Cooperon-Diffuson contributions to the probability of absorption are shown in figure \ref{fig:2} 
In the study of the crossover from large to small coherence length compared to the system size we restrict ourselves to the case of scalar waves.

\subsection{Diffusion ladders}
We use standard impurity technique while calculating correlation functions \cite{bib:abricos}.
The diffusion ladder $P(\omega,\textbf{r},\textbf{r'})$ appearing in case of multiple scattering at $\omega l/c<<1$ satisfies equation
\begin{equation}\label{dif-ladder}
(-D\nabla^{2}-i\omega)P(\omega,\textbf{r},\textbf{r'})=\delta(\textbf{r}-\textbf{r'})
\end{equation} 

Here $D=cl/3$ is light diffusion coefficient, $c$ and $l$ are velocity and mean free path of light, correspondingly.

Consider the random medium occupying the half-space $z>0$. Then the boundary condition for the ladder is given by 
$P(\omega,\textbf{r},\textbf{r}^\prime)=0$ at $z,z^\prime=0$. 
Performing the Fourier transformation over coordinates $x,y$ we obtain:
\begin{equation}
 P(\omega,z,z',\textbf{Q})=\frac{\sinh[q \min(z,z')]}{Dq}\exp[-q\max(z,z')]
\end{equation} 
where $q^2=Q^2-i\omega/D$

Considering the scattering at large angles we must distinguish between Cooperon and Diffuson propagators in the integral with four Green's functions \cite{bib:akker, bib:stephen, bib:mac}.
\begin{equation}\label{coop-def}
P_{c}(\omega,\textbf{Q})\equiv\int_{0}^{\infty}dzdz^\prime P(\omega,z,z^\prime,\textbf{Q})e^{-\frac{(z+z^\prime)}{2l}\frac{(\mu_{m}+\mu_{n})}{\mu_{n}\mu_{m}}}
\end{equation} 
and
\begin{equation}\label{diff-def}
P_{d}(\omega,\textbf{Q})\equiv\int_{0}^{\infty}dzdz^\prime P(\omega,z,z^\prime,\textbf{Q})e^{-(z/\mu_{n} l+z^\prime/\mu_{m} l)},
\end{equation} 
where $\mu_{n}$ and $\mu_{m}$ are projections of directions $-\textbf{n}$ and $\textbf{m}$ of incident and of scattered waves onto the normal to the surface.

Integrating over $z$ and $z^{\prime}$ in (\ref{coop-def}) and (\ref{diff-def}) in the limit $|q|l<1$, we obtain an expression for the Cooperon propagator:
\begin{equation}
P_{c}(\omega,\textbf{Q})= \frac{l^3}{2D}\left(\frac{2\mu_{n}\mu_{m}}{\mu_{n}+\mu_{m}}\right)^{3}\left(1-4ql\frac{\mu_{n}\mu_{m}}{\mu_{n}+\mu_{m}}\right)
\end{equation} 
and for the Diffuson propagator:
\begin{equation}
P_{d}(\omega,\textbf{Q})= \frac{l^3}{D}\frac{(\mu_{n}\mu_{m})^{2}}{\mu_{n}+\mu_{m}}\left[1-(\mu_{n}+\mu_{m})ql\right].
\end{equation} 

Equation (\ref{diff-def}) and expression (\ref{coop-def}) are valid for small momentum $ql<1$.  Considering reflection at large angle we must put $P_{c}(ql>1)\sim 0$.

\subsection{Cooperon contributions}
Cooperon contributions to the probability of absorption are shown in fig.2 a,b.
They contribute to $g^{(2)}_{1}$ when diffusion ladders couple states $k_{i}$ and
$k_{j}$ with $i=j$, and $k_{l}$ and
$k_{p}$ with $l=p$ so vertex couples states with $i\neq p$.

Let us consider first diagram, shown in Fig.2a. 
Phase factors of light incident in the direction $\textbf{n}$ and reflected in the direction $\textbf{m}$ in expression

\begin{equation}
 e^{-ik_{1}(\textbf{r}-\textbf{r}^\prime)(\textbf{n}+\textbf{m})}P_{c}(0,\textbf{r}-\textbf{r}^\prime)
e^{-ik_{2}(\textbf{R}-\textbf{R}^\prime)(\textbf{n}+\textbf{m})}P_{c}(0,\textbf{R}-\textbf{R}^\prime)
\end{equation}
must be integrated over the surface from which light is collected.

Integrating over the surface of the medium with coordinates ($\textbf{r}, \textbf{r}^\prime,\textbf{R}, \textbf{R}^\prime$) we obtain:
\begin{eqnarray}\label{coop-1f}\nonumber
&~&\int \frac{d^{2}\textbf{Q}}{(2\pi)^{2}}P_{c}(0,\textbf{Q})|F(\textbf{Q}+k_{1}(\textbf{n}+\textbf{m}))|^{2}\\
&\times&
\int \frac{d^{2}\textbf{Q}^{\prime}}{(2\pi)^{2}}P_{c}(0,\textbf{Q}^{\prime})|F(\textbf{Q}^{\prime}+k_{2}(\textbf{n}+\textbf{m}))|^{2}.
\end{eqnarray}
Where
\begin{equation}\label{ffact}
 F(\textbf{Q})=\frac{1}{S}\int_{S} d^{2}\textbf{r}\exp(i\textbf{Qr})
\end{equation}
is form factor of the surface from which light is collected.

When  $|k_{i}(\textbf{n}+\textbf{m})|$ larger than the inverse of linear dimension of area, we can calculate (\ref{coop-1f}) as:
\begin{equation}\label{finCoop1}
 P_{c}(0,k_{1}(\textbf{n}+\textbf{m}))P_{c}(0,k_{2}(\textbf{n}+\textbf{m}))S^{2}\simeq |P_{c}(0,k_{0}(\textbf{n}+\textbf{m}))S|^{2}.
\end{equation} 

$k_{0}=\Omega_{0}/c$
The contribution from the second diagram, shown in Fig. 2b, after integration of the phase factors can be written as:
\begin{equation}\label{int1}
|\int \frac{d^{2}\textbf{Q}}{(2\pi)^{2}}P_{c}(\omega_{12},\textbf{Q})F(\textbf{Q}+k_{1}\textbf{n}+k_{2}\textbf{m})F(\textbf{Q}+k_{2}\textbf{n}+k_{1}\textbf{m})|^{2}
\end{equation} 
Form-factors vary with momentum much faster than $P_{c}(\omega_{12},\textbf{Q})$, therefore the integral in (\ref{int1}) can be calculated as: 

\begin{equation}\label{finCoop2}
|P_{c}(\omega_{12},k_{0}(\textbf{n}+\textbf{m})|^{2}|F(k_{12}(\textbf{n}-\textbf{m}))|^{2}.
\end{equation} 
Here Cooperon propagators depend on the frequency $\omega_{12}=c(k_{1}-k_{2})=ck_{12}$.
The contribution from the second diagram strongly depends on the ratio between the coherence length of light and the size of the surface from which the radiation is collected. 

At  $\sqrt{S} /L_{c}<1$ and $|k_{12}|L_{c}\leq 1$ the form-factor can be approximated as  $|F(k_{12}(\textbf{n}-\textbf{m}))|\approx|F(0)|=1$. As a result, the second contribution (\ref{finCoop2}) depends only on $(\textbf{n}+\textbf{m})$ as the first one (\ref{finCoop1}). 

In the opposite case when $\sqrt{S} /L_{c}>1$ at $|k_{12}|L_{c}\sim 1$  form-factor  has maximum in direction $(\textbf{n}-\textbf{m})_{||}=0$. 
Backscattering contribution from (\ref{finCoop2}) decreases rapidly under a deviation of $\textbf{n}$ and $\textbf{m}$ from the normal to the surface.

\subsection{Diffusion contributions}

Diffusion contributions to the probability of absorption are given by two diagrams, shown in Fig 2 c, d.

Similarly to the calculation of the Cooperon contributions we integrate the phase factors over the surface and obtain for the first diagram:
\begin{equation}\label{finDif1}
| P_{d}(0,0)S |^2
\end{equation} 
Note that phase factors do not give rise to the frequency and angle dependence of the first diagram.

The second diagram can be calculated as:
\begin{equation}\label{dif2}
|P_{d}(\omega_{12},0)F(k_{12}(\textbf{n}-\textbf{m}))|^{2}
\end{equation} 

Here $k_{12}=\omega_{12}/c$. In the limit $|\omega_{12}|l/c<1$  we might neglect the momentum dependence of the diffusion ladder, therefore
$P_{d}(\omega_{12},k_{12}\textbf{n})\simeq P_{d}(\omega_{12},0)$.

Again, if detector collects the radiation from the area $S<L^{2}_{c}$ then the form factor becomes $F(k_{12}(\textbf{n}-\textbf{m}))=1$, and (\ref{dif2}) does not depend on the scattering angle. Contrary, if detector collects the radiation from large area $S>L^{2}_{c}$ then the contribution (\ref{dif2}) of the second diagram at $|k_{12}|L_{c}\sim 1$ has maximum in the direction $\textbf{m}_{||}=\textbf{n}_{||}$,  i.e.  when the angle of reflection equals the angle of incidence.

\subsection{Mixed cooper-diffusion contributions}

Diagrams that describe these contributions are shown in Fig. 2e,f.
After the integration over the surface of the medium the contribution of mixed diagrams, shown in Fig. 2e, results in: 
\begin{equation}
\int\frac{d^{2}\textbf{Q}}{(2\pi)^{2}} [P_{c}(0,\textbf{Q})F^{2}(\textbf{Q}+k_{1}(\textbf{m}+\textbf{n}))+(k_{1}\rightarrow k_{2})]~P_{d}(0,0)S.
\end{equation} 
We then perform the integration over the momentum and obtain:
\begin{equation}\label{mix1}
2 P_{c}(0,(k_{0}(\textbf{m}+\textbf{n}))P_{d}(0,0)S^{2}.
\end{equation} 

The diagram shown in Fig. 2f after the integration of phase factors over the surface of the medium yields:
\begin{eqnarray}
&2&\mathrm{Re} \int\frac{d^{2}\textbf{Q}}{(2\pi)^{2}} P_{c}(\omega_{12},\textbf{Q})F(\textbf{Q}+k_{2}\textbf{m}+k_{1}\textbf{n})\\ \nonumber
&\times& F(\textbf{Q}+k_{1}\textbf{m}+k_{2}\textbf{n})
P_{d}(-\omega_{12},k_{12}\textbf{n})F(k_{12}(\textbf{n}-\textbf{m})).
\end{eqnarray}

Again, the integration over momentum gives:
\begin{equation}\label{mix2}
2\mathrm{Re} P_{c}(\omega_{12},k_{0}(\textbf{n}+\textbf{m}))P_{d}(-\omega_{12},0)|F(k_{12}(\textbf{n}-\textbf{m}))|^{2}.
\end{equation} 

\subsection{Sum of all contributions}

At $l \ll L_{c}, \sqrt{S}$ we can neglect the frequency  dependence of $P_{d}$ and $P_{c}$. 

Collecting all Cooperon (\ref{finCoop1}, \ref{finCoop2}), Diffuson contributions (\ref{finDif1}, \ref{dif2}) and
mixed contributions (\ref{mix1}, \ref{mix2}) we obtain in the limits $L_{c}, \sqrt{S}>>k_{0}|\textbf{n}_{||}+\textbf{m}_{||}|$:

\begin{eqnarray}\label{total}
&~&\Sigma(\omega_{12})\equiv 
[P_{c}(0,k_{0}(\textbf{n}+\textbf{m}))+P_{d}(0,0)]^{2}\times \\\nonumber
&\times &(1+|F(k_{12}(\textbf{n}-\textbf{m}))|^{2})
\end{eqnarray}

Diagrams that determine $g^{(2)}_{0}$ can be obtained from that, shown in Fig. 1, by interchanging
Green's functions in such a way that there is no change of state of light at the vertex, i.e. diffusion ladders couple state 
$i=p$ to $j=l$ in figure 1.
The sum of such diagrams is equal to (\ref{total}).

\subsection{Integrating over frequencies}
To obtain $g^{(2)}_{1}$  we  integrate (\ref{total}) over frequencies. $\Sigma(\omega_{12})$ depends of frequencies difference, therefore
\begin{eqnarray}\label{int-freq}
&~&\int d\omega_{1}d\omega_{2}\nu (\omega_{1})\nu(\omega_{2})\delta(\Omega-\omega_{12})\Sigma(\omega_{12})\\\nonumber
&\simeq& \sqrt{\frac{\pi}{2\sigma}} N^{2}_{0}\Sigma(\Omega)\exp(-\sigma\Omega^2/2),
\end{eqnarray}

As normalization constant we choose square of average intensity of light without interference correction, 
\begin{equation}
\langle G^{(1)} \rangle^{2}=(SP_{d}(0,0)N_{0}\int d\omega \exp(-\sigma\omega^{2}))^{2} 
\end{equation}

Therefore

\begin{eqnarray}\label{result1}
g^{(2)}_{1}=\sqrt{\sigma/2\pi}\exp(-\sigma\Omega^{2}/2)(1+\frac{P_{c}(0,k_{0}(\textbf{n}+\textbf{m}))}{P_{d}(0,0)})^{2}\times\\\nonumber
\times (1+|F(\Omega(\textbf{n}-\textbf{m})/c)|^{2})
\end{eqnarray}

Integrating (\ref{result1}) over $\Omega$ we obtain

\begin{equation}\label{result2}
g^{(2)}_{0}
=(1+\frac{P_{c}(0,k_{0}(\textbf{n}+\textbf{m}))}{P_{d}(0,0)})^{2}(1+\Phi(\textbf{n}-\textbf{m}))
\end{equation}

Function $\Phi(\mathbf{ p})$ is determined by the form-factor (\ref{ffact}) of the surface, from which the light is collected, as
\begin{equation}
 \Phi(\textbf{p})=\sqrt{\frac{\sigma}{2\pi}}\int_{-\infty}^{\infty}d\omega
\left| F(\omega\textbf{p}/c)\right|^{2} e^{-\sigma\omega^2/2}. 
\end{equation}

By definition, functions  $ F(\omega\textbf{p}/c)$ and $\Phi(\mathbf{ p})$ depend only on the components of the vector $\textbf{p}$ parallel to the surface.

\section{Discussion}
Functions $|F(\Omega(\textbf{n}-\textbf{m})/c)|$ and $\Phi(\textbf{n}-\textbf{m})$  when observed at angles equal to the angle of incidence  $(\textbf{n}-\textbf{m})|_{\|}=0$ have maximum which is equal to one. At $L_{c}<<\sqrt{S}$, $\Omega L_{c}/c\geq 1$ the maximum resembles a peak of $|\textbf{n}_{\|}-\textbf{m}_{\|}|\sim \sqrt{S}/L_{c}$ wide.
$\textbf{n}_{||}$ and $\textbf{m}_{||}$ are the components of direction of light parallel to the surface. This behavior is reminiscent of "memory effect"  \cite{bib:feng} in transmission.

Factor $(1+\frac{P_{c}(0,k_{0}(\textbf{n}+\textbf{m}))}{P_{d}(0,0)})^{2}$ describes backscattering triangular like peak at angles $|\textbf{n}+\textbf{m}|\leq (k_{0}l)^{-1}<<1$.

Two cases might be distinguised. The angle of incidence is zero and of the order of one.

In first case peaks are superimposed.

\begin{figure}[t]  \centering
\includegraphics[width=8cm] {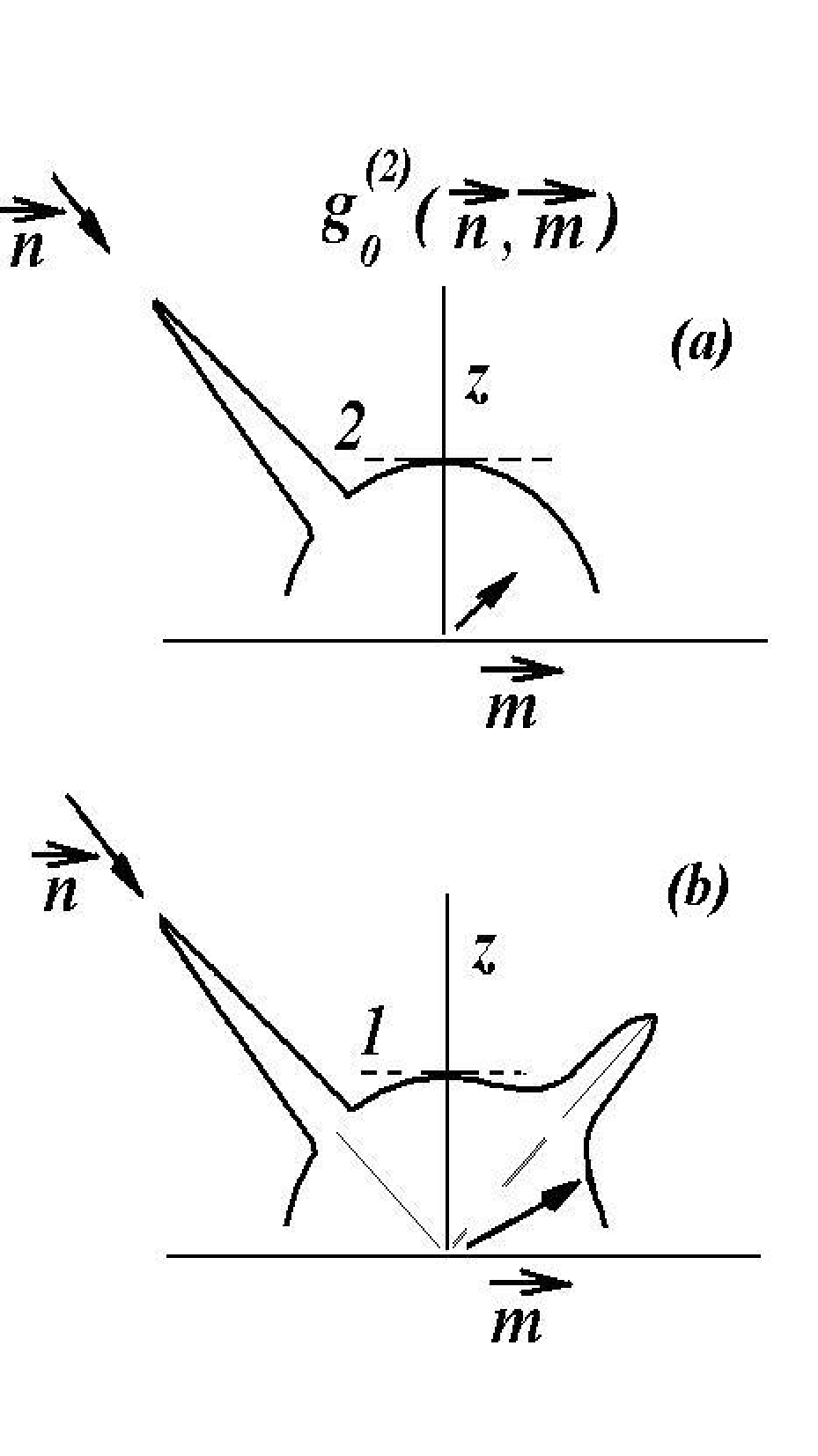}
\caption{Schematic dependence of $ g^{(2)}_{0}(\textbf{n},\textbf{m})$ of the scattering angle $\textbf{m}$ at different value of $L_{c}/\sqrt{S}$. 
Figure (a) corresponds to large ratio. 
Figure (b) corresponds to case $\sqrt{S}>L_{c}$}. Here the coherent contribution is suppressed and the peak remains in forward scattering direction.\label{fig:3}
\end{figure}

At large incidence angle we obtain two peaks in reflection. One in backscattering direction, and second in the forward-scattering direction. The amplitude of second peak is of order of one at small coherence length, as it is shown on figure \ref{fig:3}.

Note that the appearance of this peak is associated with suppression of the background  in the case of incoherent light.

\subsection{Time dependence}

Peak in (\ref{result1}) develops at relatively high frequency, i.e. at short time. It is interesting to consider time evolution, related with this term.

\begin{eqnarray}
\int\frac{d\Omega}{2\pi}\exp(i\Omega t-\sigma\Omega^{2}/2)(1+|F(\Omega \delta\textbf{n}/c)|^{2})=\nonumber\\
=\frac{\exp(-t^{2}/2\sigma)}{\sqrt{2\pi \sigma}}+\int\frac{d^{2}R_{1}d^{2}R_{2}}{\sqrt{2\pi \sigma}S^{2}}\exp(-(t+\delta \textbf{n}\textbf{R}_{12})^{2}/2\sigma)\nonumber
\end{eqnarray}

For square $|x|,|y|<L/2$ and $\delta \textbf{n}||x $ second term is

\begin{eqnarray}\label{time-evol}
2\exp(-(ct/L_{c})^{2}/2)\int^{1}_{0}dp(1-p)\cosh(\frac{\delta nLct}{L_{c}^{2}}p)\times\\
\times\exp(-(\delta nLp/L_{c})^{2}/2)\nonumber
\end{eqnarray}

At the beginning, when $ct/L_{c}\leq 1$ exponential term in (\ref{time-evol}) wins. Expression 
(\ref{time-evol}) has maximum at small $\delta n$. At $L_{c}<L$ it decreases $\sim L_{c}/\delta n L$. Peak at small $\delta n$ disappears at larger time $tc>L_{c}$.

\section{Conclusion}
Let us consider how the divergence of the incident beam limits the proposed interference picture. Let we have two incident beams 
characterized by $k_{1}, \textbf{n}_{1}$ and $k_{2}, \textbf{n}_{2}$.  The form-factor in this case is
$F(k_{12}(\textbf{m}-(\textbf{n}_{1}+\textbf{n}_{2})/2)-(k_{1}+k_{2})(\textbf{n}_{1}-\textbf{n}_{2})/2)$.
If $k_{0}>>|k_{12}|$ the most important limitation is associated with the second term of the argument of the form-factor.
Therefore for observation of the peak in absorption probability at forward scattering direction divergence of incident beam must be 
small, so that condition $k_{0}|\textbf{n}_{1}-\textbf{n}_{2}|\sqrt{S}<1$ is satisfied.

Note that recent experiments in Ref. \cite{bib:scalia} were performed 
with pseudo thermal light which has coherence time $\sim 10^{-6}$ sec and coherence length of hundreds of meters.
Comparing to system size this is too much for observation of changing of reflection structure at angle equal to angle of incidence.

To conclude, we calculate the probability of absorption of two photons reflected from the random medium as a function of the reflection 
angle.
We show that result depends on the ratio between the size of the medium and coherence length of light. We predict a peak in 
absorption probability when the angle of reflection is equal to the angle of incidence in case when coherence length is smaller than the system size. 

We are grateful for the financial support of RFFI under Grant No.
12-02-00300-A.

\end{document}